\documentclass[twocolumn,prb,aps,floatfix,amssymb]{revtex4}
\usepackage{graphicx}
\usepackage{multirow}
\usepackage{color}
\usepackage{amsmath}
\usepackage{amssymb}

\usepackage{chngcntr}
\counterwithin*{equation}{section}
\counterwithin*{equation}{subsection}

\begin{document}

\title{Effects of interedge scattering on the Wigner crystallization in graphene nanoribbons}

\author{Mohsen Modarresi$^{1,2}$ and A. D. G\"u\c{c}l\"u$^1$}
\affiliation{$^1$Department of Physics, Izmir Institute of Technology, IZTECH, TR35430, Izmir, Turkey.}
\affiliation{$^2$Department of Physics, Ferdowsi University of Mashhad, Mashhad, Iran.}

\begin{abstract}
We investigate the effects of coupling between the two zigzag edges of graphene nanoribbons on the Wigner crystallization of electrons and holes using a combination of tight-binding, mean field Hubbard and many-body configuration interaction methods. We show that the thickness of the nanoribbon plays a
crucial role in the formation of Wigner crystal. For ribbon widths smaller than 16 \mbox{\AA}, increased kinetic energy overcomes the long-range Coulomb repulsion and suppresses the Wigner crystallization. For wider ribbons up to 38 \mbox{\AA} wide, strong Wigner localization is observed for even number of electrons, revealing an even-odd effect also found in Coulomb blockade addition spectrum. Interedge correlations are found to be strong enough to allow simultaneous crystallization on both edges, although an applied electric field can decouple the two edges. Finally, we show that Wigner crystallization can also occurs for holes, albeit weaker than for electrons.
\end{abstract}

\maketitle 

\section{Introduction}

As the density of an electron gas is lowered, Coulomb energy can overcome the kinetic energy which causes the electrons to localize at their classical position in order to minimize the electron-electron interactions. This process is known as the Wigner crystallization \cite{Wigner,Tanatar,Guclu4,Guclu6}, which is expected to have important implications on electronic transport properties \cite{Steinberg,Auslaender,Guclu3}. The electron crystal phase was first observed on the surface of liquid helium \cite{Grimes}. Later, the metal-insulator transition in GaAs/AlGaAs heterostructures was also attributed to the Wigner crystallization \cite{Jongsoo}. In the case of bulk graphene, the linear dispersion of Dirac fermions is expected to prevent formation of Wigner crystal 
 \cite{guclu,Solyom}, unless a strong magnetic field is applied\cite{Joglekar} or finite size effects are present\cite{guclu,add1,add2}. 

In one-dimensional electron gas, even a very weak long-range force is expected to lead to a one-dimensional Wigner crystal\cite{Schulz}, although quantum fluctuations may prevent the formation of the electronic localization \cite{Solyom}. The first one-dimensional transport measurements was performed in GaAs heterostructures point contacts \cite{Van}. Later, localization in one-dimensional GaAs/AlGaAs heterostructures was observed by measuring the tunneling conductance \cite{Steinberg,Auslaender} and was investigated theoretically in inhomogeneous one-dimensional systems \cite{Guclu3,Mueller}. On the other hand, the low-temperature single-electron transport spectroscopy was used to show the formation of one-dimensional Wigner crystal in carbon nanotubes \cite{Vikram}. Also, Wigner molecules were experimentally observed in ultra-clean carbon nanotubes \cite{Pecker}.

The zigzag graphene nanoribbon (ZGNR) is a one dimensional strip of graphene with zigzag edges. The presence of a highly degenerate band of zigzag edge states is expected to give rise to unusual magnetic properties\cite{Kang, DJiang,guclu2, Wimmer,Yazyev,Yang,Jung1,Jung2,Soriano,Feldner,Ozdemir}, similar to graphene quantum dots with zigzag edges\cite{Potasz1,Potasz2,Devrimbook,Guclu5,Akhmerov1}. Although there is no direct experimental evidence of edge magnetization in such structures, recent experimental works \cite{Rao,Magda} indicate possible ferromagnetic-antiferromagnetic phase transition in ZGNRs, which may be related to disorder effects \cite{Ozdemir}. Graphene nanoribbons were synthesized using various experimental methods \cite{Jiao,Talyzin,Banhart,Cai}.
Moreover, recently highly clean ZGNRs with well-defined size were synthesized by surface assisted polymerization of a specific monomer \cite{Ruffieux}. 

Another interesting property of zigzag edges in graphene nanostructures is the possibility of Wigner localization predicted theoretically at relatively high electronic densities in a single zigzag edge \cite{guclu}. However, in ZGNRs, the effects of interedge correlations due to electronic interactions remains unknown. In this work, we use a combination of tight-binding, mean field Hubbard and many-body configuration interaction methods in order to study the effects of interedge correlations on Wigner crystallization of electrons and holes in ZGNRs as a function of size and electronic density in the edge states. We show that the thickness of the nanoribbon plays a crucial role in the Wigner crystallization. In particular, while for wide ribbons up to 38 \mbox{\AA} strong Wigner localization is observed on both edges simultaneously, increased kinetic energy suppresses the crystallization in ribbons thinner than 16 \mbox{\AA} wide. Moreover, the analysis of Coulomb blockade addition spectrum shows the formation of meta-stable ground states for even number of electrons, leading to oscillatory behaviour consistent with classical calculations.

\section{Model and Method}

In order to include electronic correlation effects within the electrons occupying the edge states, we start with the tight-binding Hamiltonian expanded in the localized atomic basis set of $p_z$ atomic orbitals. The nearest-neighbor and next-nearest-neighbor hopping parameters between two carbon atoms are set to -2.7 $eV$ and -0.1 $eV$, respectively. The low energy states around the Fermi level are well localized at the zigzag edges of the ribbon. Once those edge states are identified (explained below), we proceed with the meanfield Hartree-Fock approach to include the effect of electron-electron interactions in the bulk states as discussed in the appendix. Once the self-consistent mean field problem is solved, the obtained eigenstates can then be used as a basis set to expand the many-body Hamiltonian

\begin{eqnarray}
H= \sum_{p\sigma} E_{p\sigma} c_{p\sigma}^{\dagger }c_{p\sigma}+\frac{1}{2}\sum_{pqrs}\sum_{\sigma \sigma^{'}}\left \langle pq \left | V \right | rs\right \rangle c_{p\sigma}^{\dagger } c_{q\sigma^{'}}^{\dagger }c_{r\sigma^{'}}c_{s\sigma}
\end{eqnarray}

where $E_{p\sigma}$ and $c_{p\sigma}^{\dagger }(c_{p\sigma})$ are the energy levels and the creation (annihilation) operator for electron with spin $\sigma$ in the $p$-th state of the mean field Hamiltonian.  The two-body interaction terms consist of the simultaneous scattering of electron in $p\rightarrow s$ and $q\rightarrow r$ states with the matrix elements $\left \langle pq \left | V \right | rs\right \rangle$, expressed in terms of two-body localized $p_z$ orbital scattering matrix. In Rydberg units, the two-body matrix elements are calculated as,  

\begin{multline}
\left \langle pq \left | V \right | rs\right \rangle = \int \int {d\bf{r_1}} {d\bf{r_2}} \phi_{p}^{*}(\bf{r_1}) \phi_{q}^{*}(\bf{r_2})  \cr \frac{2}{\kappa \mid \bf{r_1}-\bf{r_2} \mid}\phi_{r}(\bf{r_2}) \phi_{s}(\bf{r_1})
\end{multline}

where $\phi_{p}^{*}(\bf{r})$ is the $p$-th tight binding localized edge state at $\bf{r}$ and $\kappa$ is the dielectric constant which is set to 6.
The interaction strength between nearest and next nearest atoms are tabulated (see Table I in the appendix) and for other atomic sites we used the screened Coulomb term. For the localized edge states around the Fermi level we use the configuration interaction (CI) method to solve the many-body Hamiltonian and obtain the many-body eigenstates of ZGNR. For CI calculations, the Hilbert space basis is constructed by consecutively applying all possible combination of creation operator to the vacuum state. The CI method is an accurate way to solve the many-body Schrodinger equation by directly solving the many-body Hamiltonian. It includes all correlation effect missing in the density functional and Hartree-Fock theories \cite{Devrimbook}. In our calculations, we consider ZGNRs with and without periodic boundary condition \cite{guclu} with $W$ atoms wide and $L$ atoms long as depicted in Fig.\ref{FIG1}a. For plotted ribbon in Fig.\ref{FIG1}a, $L$=52 and $W$=28 which corresponds to approximately 65 \mbox{\AA} long and 28 \mbox{\AA} wide. 

\section{Results and Discussion}

\begin{figure}[!t]
\includegraphics[width=0.5\textwidth,clip]{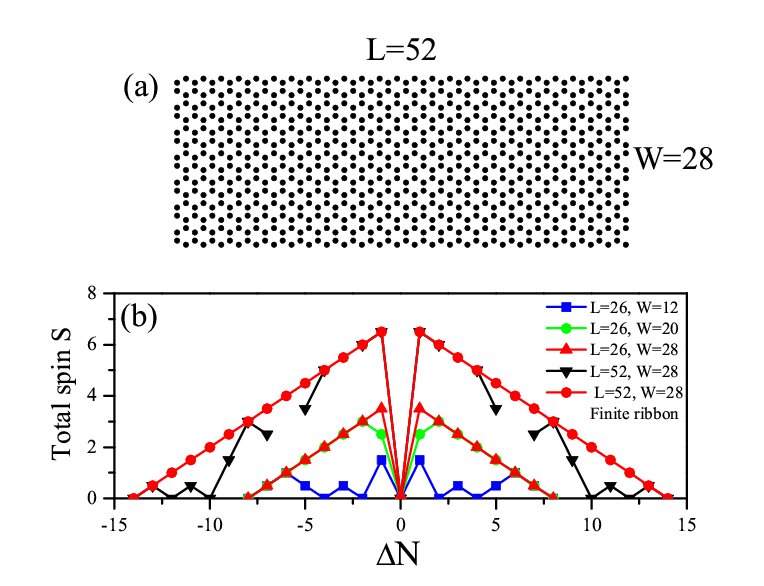}
\caption{(a) The atomic structure of ZGNR for $L$=52 and $W$=28, (b) Variation of total spin $S$ with number of electrons for ZGNR for different width and length.}
\label{FIG1}
\end{figure}

In our CI calculations, we consider a total of $N=N^{up}+N^{down}$ spin up and down electrons occupying the zigzag edge states. The edge states are not quite flat and possess different localization extent. One can identify the edge states by analyzing the wave functions from tight binding Hamiltonian. For a periodic ring, the edge states can be identified through their Fourier index $k$ between $L/6$ and $L/3$ \cite{guclu2} which gives as estimation of their number. In practice, we also check the convergence of CI results as we increase the number of edge states included in the calculations. The many-body Hamiltonian can be conveniently solved in the subspace of the total spin component $S_z=\frac{1}{2}\left [ N^{up}-N^{down} \right]$. Once the eigenenergies are calculated, the net total spin $S$ of the ground state can be deduced. In our calculations, the size of the largest Hamiltonian matrix diagonalized using the Lanczos algorithm was $1093092\times1093092$. 

The variation of total spin $S$ with the charge doping with respect to the charge neutral system, $\Delta N=N-N^{neutral}$, is plotted in Fig.\ref{FIG1}b. We note that the total number of electrons in the ZGNR is the summation of edge electrons $N$ and number of electrons in the bulk states. For half-filling (charge neutrality), the total spin is $S$=0 and we observe antiferromagnetic configuration of spins on opposite zigzag edges which was reported by mean field Hubbard and density-functional-theories \cite{Eduardo,Jung,Jiang}. For wider ribbons and away from charge neutrality, the total spin of ground states reaches to the maximum possible value $S^{max}=\frac{1}{2}N$ giving a ferromagnetic coupling between the two edges. However, for the periodic ribbon, if we increase the ribbon length to $L$=52, total spin $S$ is not maximized anymore and spin oscillations are observed, indicating that magnetic properties are sensitive to both the length and the width of the ribbon. We note that by increasing the ribbon length, the number of edge states also increases, making the computational calculations more difficult. The missing points in Fig.\ref{FIG1}b are due to computational limitations for large Hamiltonian matrices. In contrast with the periodic ribbon, the ground state for the finite ribbon away from the charge neutrality is completely spin polarized $S^{max}$. We note that in the case of periodic ribbon, the number of unit cells along the ribbon length may break or restore the inversion symmetry which may have additional effects on the numerical results requiring further investigation. Here, we fixed the length of ribbon to 26 and 52 atomic sites in all calculations.

\begin{figure}[!t]
\centering
\includegraphics[width=\linewidth]{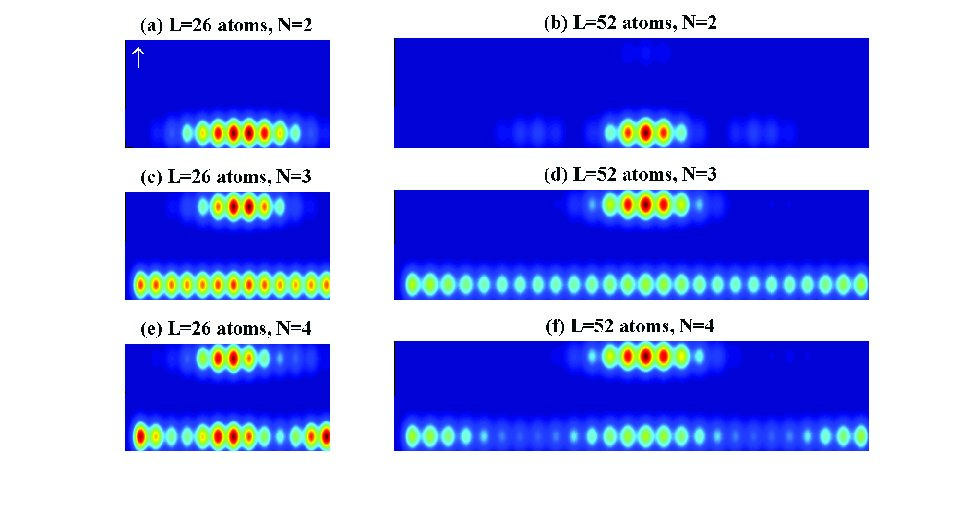}
\caption{Pair-correlation function for periodic ZGNRs with $W$=28 atoms, (a,c,e) $L$=26 and (b,d,f) $L$=52 and $N$=2,3 and 4 electrons. The fixed electron is located at the upper left corner of the ribbon in all cases, indicated by a white arrow in (a). For clarity, part of the empty bulk region of the ribbon is cropped from the plot.}
\label{FIG2}
\end{figure}

We now investigate the pair-correlation function (PCF), i.e. the probability of finding an electron with spin $\sigma^{'}$ at a position $r^{'}$, provided that another electron with spin $\sigma$ is present at the position $r$, calculated using the ground state expectation value $P_{\sigma \sigma^{'}}({\bf r} , {\bf r}^{'})=\langle n_{\sigma}({\bf r}) n_{\sigma^{'}}({\bf r}^{'})\rangle$, where $n$ is the density operator. Figure \ref{FIG2} shows the PCF for $W$=28 and $L$=26 and 52 atoms for a periodic ZGNR. For the numerical calculation of PCFs, the position ${\bf r}$ of an electron with spin up is fixed at the top left corner of the ribbon, shown with a white arrow in Fig.\ref{FIG2}a. The total PCF is the sum over spin up and down probabilities, $P=P_{\uparrow\uparrow}+P_{\uparrow\downarrow}$. Fig.\ref{FIG2} is the counter plot of PCF as a function of ${\bf r}^{'}$. 
 
For clarity, part of the empty bulk region of the ribbon was cropped from the figure. For $N$=2, one electron is localized in the center of the lower edge to form a zigzag configuration with the reference electron at the upper left corner. Clearly, the system is in the Wigner crystal regime. When we add one more electron, although the upper edge seems to remain crystallized, we observe a nearly homogeneous electronic distribution on the lower edge. One might think that the Wigner crystal is destroyed due to the increased electronic density. However, the suppression of Wigner crystal here is mainly due to the fact that an odd number of electrons must be shared between the two edges. Indeed, the ground state is doubly degenerate with a linear combination of two possible classical configurations. For plotting purposes, we have taken the average PCF of the two degenerate ground states. As we increase the number of electrons to $N$=4, each edge now hosts two electrons that are localized at the classically predicted positions. The electronic density of up and down edges are coupled together via the long-range Coulomb interaction to form the Wigner crystal. Note that, unlike for the $N$=2 that exhibits a zigzag Wigner configuration, the $N$=4 system has a symmetric configuration (upper and lower edge electrons are lined up) which is consistent with our classical calculations for the same system (not shown).

\begin{figure}[!t]
\includegraphics[width=\linewidth]{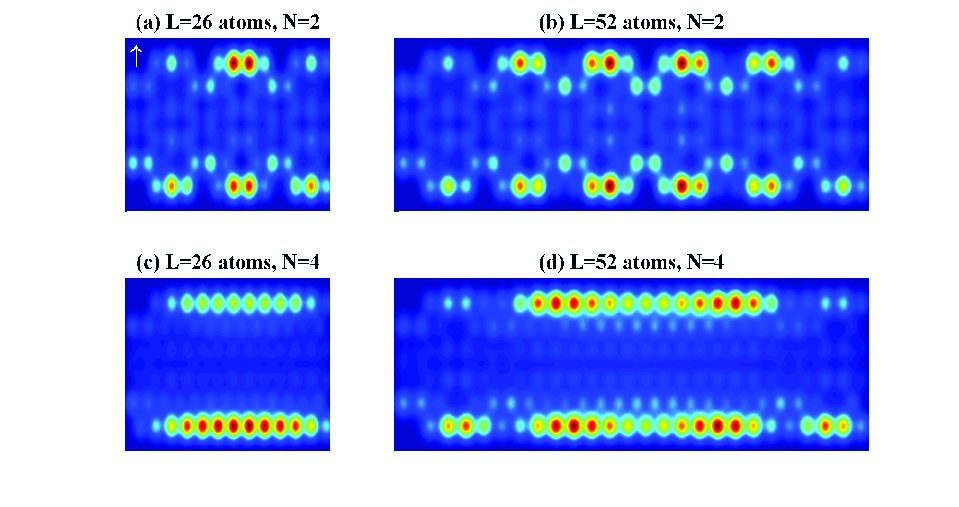}
\caption{Pair-correlation function for narrow periodic ZGNRs $W$=12 with (a,c) $L$=26 and (b,d) $L$=52. The fixed electron is located at the upper left corner of the ribbon in all cases, indicated by a white arrow in (a).}
\label{FIG3}
\end{figure}

In Fig.\ref{FIG3}, we focus on narrower periodic nanoribbons with $W$=12 and two different lengths, $L$=26 (left panels) and 52 (right panels), similar in size to the recently produced experimental ribbons in Ref.[\onlinecite{Ruffieux}]. The top and bottom panels show the PCFs corresponding to $N$=2 and 4 electrons, respectively. For these structures, although charge density oscillations are present, no clear signature of Wigner crystallization is observed. In fact, due to the narrow nature of the ribbon, the opposite edge states have a high spatial overlap, lifting the degeneracy of the zero-energy band. As a result, the relative kinetic energy of the electrons occupying the edge states increases, suppressing the effect of long-range electron-electron interactions. 

\begin{figure}[!t]
\includegraphics[width=\linewidth]{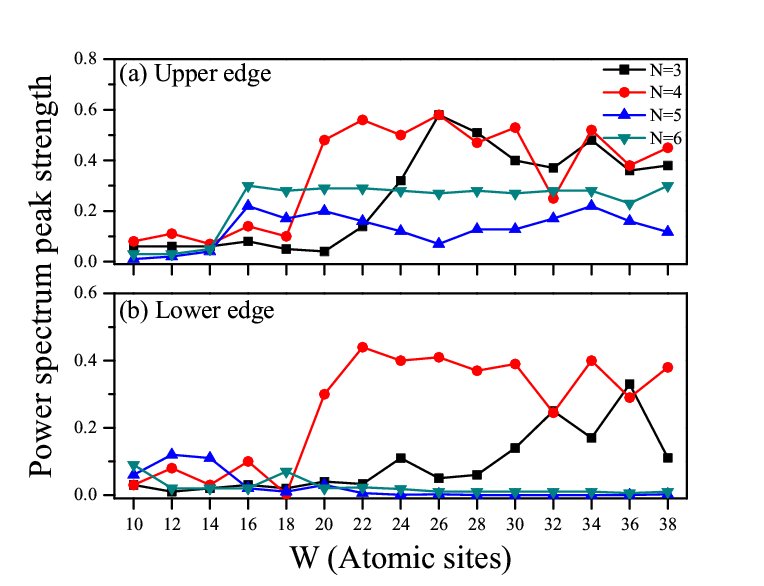}
\caption{Power spectrum peak strength as a function of ribbon width for a ribbon with L=52 and (a) upper and (b) lower edge.}
\label{FIGWW}
\end{figure}

In order to investigate the effect of ribbon width on the electronic localization further, 
we calculated the power spectrum, i.e. the Fourier transform of the PCF, which is a more useful way to quantify the degree of localization in periodic systems\cite{guclu,Guclu4}.
In the following, we consider the $S^{max}$ as the ground state in our calculations. The alignment of all spin in the system is achievable in the presence of an in-plane external magnetic field. A discussion of the effect of spin on the Wigner localization at zigzag edges can be found in Ref.[\onlinecite{guclu}]. A strong peak maximum at the Fourier component $k=N_{edge}$ in the power spectrum indicates the localization of $N_{edge}$ electrons on the edge. Figure \ref{FIGWW} shows the power spectrum peak strength corresponding to upper and lower edges as a function of ribbon width $W$ for different electron numbers $N$. We observe small oscillations as a function of $W$, presumably due to the change in the symmetry of the ribbon as the $W$ is increased by two (i.e. by an extra zigzag chain) \cite{symetry}. More strikingly, there exists a critical value of the width, $W=16$, above which the upper edge electrons containing the fixed electron become strongly localized. In the lower edge, localization is weaker but still present especially for $N=3$ and 4 and starts at a higher critical value of $W=20$, indicating strong interedge correlations. The specific value of critical width for Wigner crystallization presumably depends on the dielectric constant which may differ due to the presence of substrate. Also, we neglect inhomogeneities in the substrate and imperfections in the graphene lattice \cite{Ozdemir}, which may affect the critical width. Further experimental and theoretical work are needed for an accurate determination of the critical width for graphene nanoribbons under different experimental conditions.

\begin{figure}[!t]
\includegraphics[width=0.4\textwidth,clip]{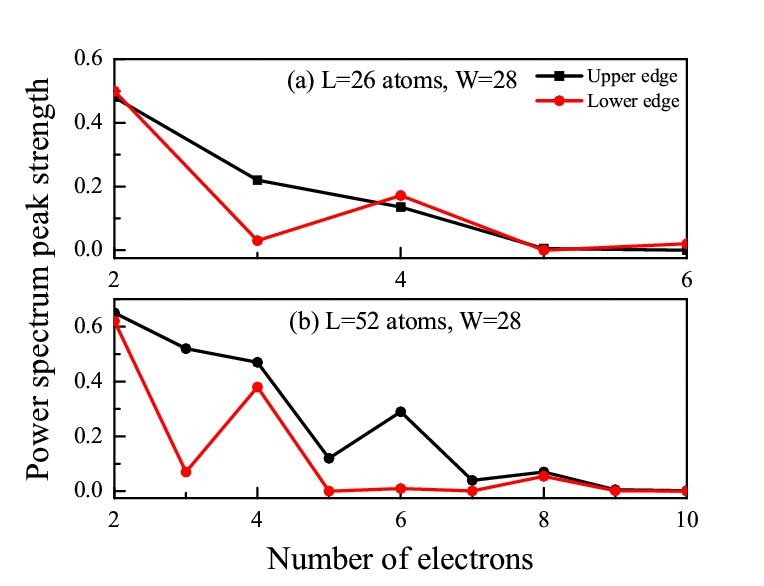}
\caption{Power spectrum peak strength in upper and lower edges as a function of $N$ for periodic ZGNR with $W$=28, (a) $L$=26 and (b) $L$=52.}
\label{FIG5}
\end{figure}

Figure \ref{FIG5} shows the strength of power spectrum peak maxima for $L$=26, 52 and $W$=28 as a function of electron number $N$. Consistent with our previous discussion on the even-odd effect, the power spectrum oscillates with number of electrons. For even number of electrons the Wigner crystal is formed on both edges for up to eight electrons, but the localization is more robust on the upper edge where the reference electron is fixed. We observed the Wigner localization for electronic densities up to 1.2 $nm^{-1}$ which is much higher than the critical density for a one-dimensional electron gas \cite{Steinberg,Guclu3}.

\begin{figure}[!t]
\includegraphics[width=\linewidth]{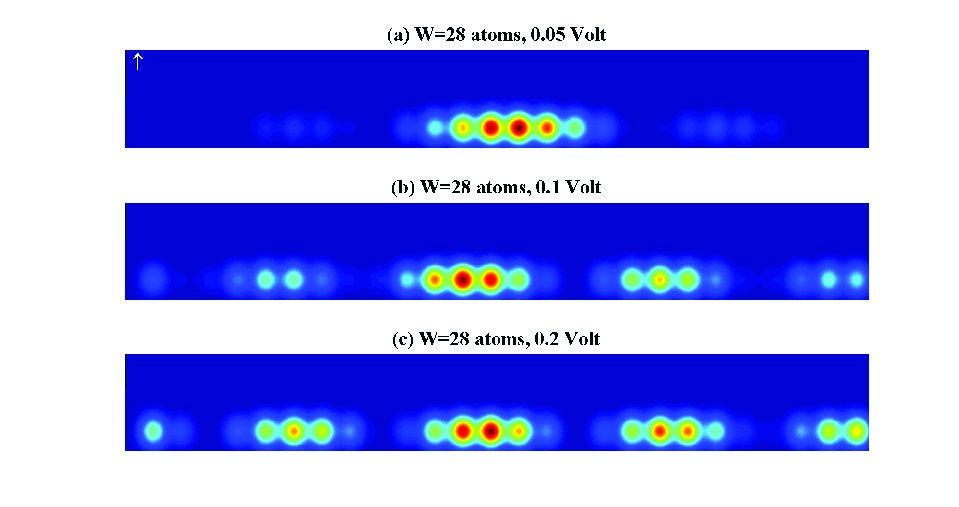}
\caption{The PCF for ribbon with L=52, W=28 atoms and N=2 electrons for different applied voltages.}
\label{FIGWV}
\end{figure}

It is possible to control the interedge correlations described above by applying an external electric field through the ribbon, creating a potential difference $V$ between the two edges. Figure \ref{FIGWV} shows the effect of the potential difference $V$ on the PCF for $N=2$. By increasing $V$, localization on the lower edge becomes suppressed. Indeed, the potential difference decouples the edge states energetically, destroying the interedge correlations. Moreover, for odd $N$ it is also possible to control the relative number of electrons on each edge preventing the edges from sharing an electron (not shown).

\begin{figure}[!t]
\includegraphics[width=\linewidth]{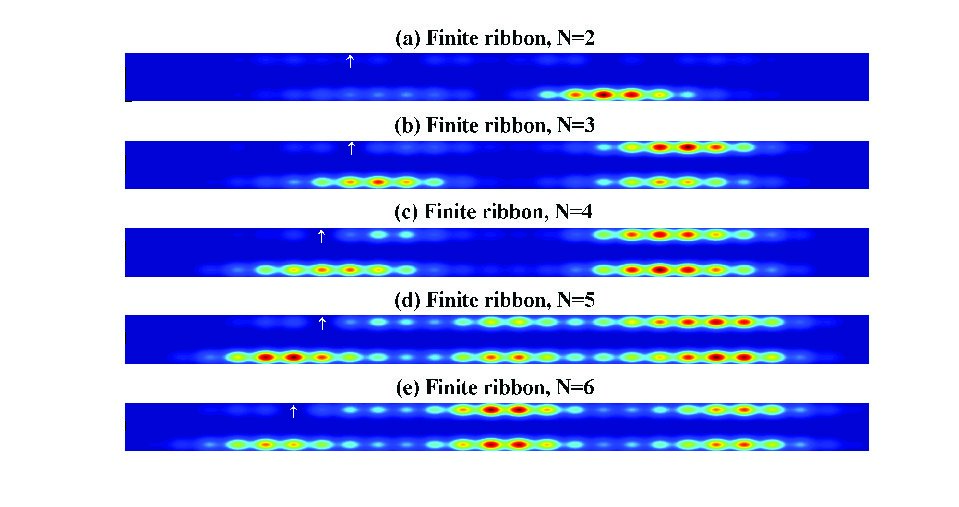}
\caption{Pair-correlation function for a finite ZGNR with $L$=52 and $W$=28 atoms and up to 6 electrons. The white arrows show the position of fixed electron which is chosen to be at a density maximum position.}
\label{FIG4}
\end{figure}

We now investigate the effect of boundary conditions on the Wigner crystallization. Figure \ref{FIG4} shows the pair-correlation function for a finite ribbon instead of periodic, with $L$=52 and $W$=28 atoms. We consider up to 6 electrons which are localized at the zigzag edges. The fixed electron is set in the charge density maximum and indicated with white arrows. Due to the existence of armchair edges at the left and right sides of the ribbon, the electrons are pushed towards the middle of ZGNR due to quantum confinement effects unlike what would happen in a classical system. For even number of electrons, Wigner localization is again clearly observed in all cases. But for odd number of electrons, although charge density oscillations are present they do not match the classical configurations nor the expected number of peaks. Hence, as for periodic ZGNRs, the even-odd effect plays an important role in the formation of Wigner crystal in a finite ZGNR.

\begin{figure}[!t]
\includegraphics[width=0.4\textwidth,clip]{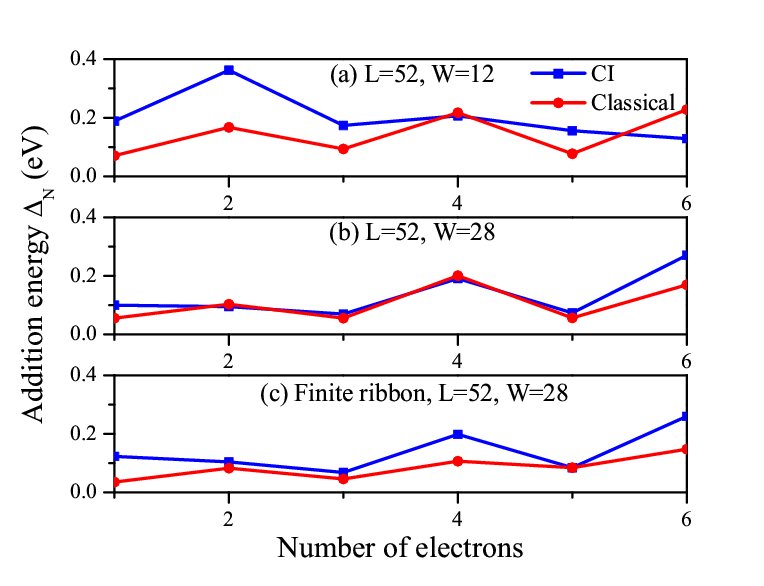}
\caption{The addition energy spectra in the classical and CI models for $L$=52, (a) periodic ribbon with $W$=12 and (b) periodic ribbon with $W$=28 atoms (c) Finite ribbon with $W$=28.}
\label{FIG6}
\end{figure}

A possible way to probe Wigner crystallization and strong interaction effects in nanostructures is Coulomb blockade spectroscopy \cite{Klein,Guclu4}. In confined structures, the large electrostatic charging energy forces the system to have a fixed number of electrons, preventing the current flow. By applying a gate potential it is possible to lift this Coulomb blockade. Thus, the conductance through the system as a function of gate voltage gives a series of sharp peaks with spacing between them proportional to the second difference between of the ground state energy with respect to the electron number $N$. In terms of the chemical potential $\mu$, the addition energy for the $N$-th peak is given by $\Delta _{N}=\mu _{N}-\mu _{N-1}$. Figure \ref{FIG6} shows the addition energy spectrum for ZGNR with $L$=52 and $W$=12,28 obtained from CI ground state energies. Note that within the non-interacting tight-binding model the addition energies would be very small due to the nearly degenerate edge states. However, including interaction effects and due to the presence of two edges of the ribbon, the addition energies are different for even and odd number of electrons, which is reflected in Fig.\ref{FIG6}. In the Fig.\ref{FIG6} we also compare the addition spectrum in the classical and CI models. In the classical model we consider localized electrons on the atomic sites with direct Coulomb interaction. All different possible electron configurations were examined to minimize the total energy to find the ground state of classical model. The simple classical calculations predict even-odd effect based oscillations in the addition energy with number of electrons. In the CI calculations the same behaviour is observed for $W$=28 but for the $W$=12 the CI model does not follow the classical results. For $W$=28 the electronic density is well explained in the classical limit which confirms the formation of a Wigner crystal for wide ZGNR. We note that the agreement between the CI and classical results is better for the periodic ribbon than for the finite ribbon. This is due to quantum confinement effects present in the finite ZGNR which pushes the electrons towards the middle of the zigzag edges resulting in a electronic configuration different from classically predicted configuration.  

\begin{figure}[!t]
\includegraphics[width=0.5\textwidth,clip]{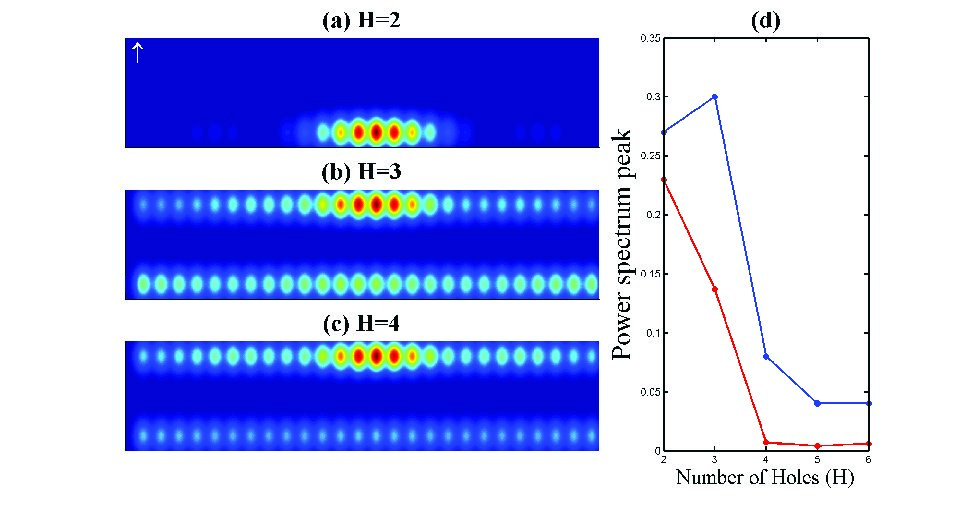}
\caption{Pair-correlation function for holes in periodic ZGNR with $L$=52, $W$=28 and(a) $H$=2, (b) $H$=3 (c) $H$=4 holes (d) The hole power spectrum peak height for the upper (blue) and lower (red) edges of ZGNR.}
\label{FIG7}
\end{figure}

An interesting question is whether edge holes in a zigzag ribbon (i.e. absence of electrons in fully occupied edge states) can Wigner crystallize. The possibility of hole localization is investigated in Fig.\ref{FIG7}. The PCF for $H$=2,3, and 4 holes for the ribbon with $L$=52 and $W$=28 is plotted in Fig.\ref{FIG7}a-c, where the reference hole is fixed at the top left corner of the system as before. 
Note that according to Fig.\ref{FIG1}b the total spin has electron-hole symmetry with respect to the half-filling of the edge states, thus hole crystallization is not unexpected. From the other point of view the electron-hole system is not symmetric due to the next-nearest neighbor hopping terms in the original tight binding Hamiltonian. Indeed, we observe a Wigner crystal for $H$=2, and a delocalized opposite edge charge distribution for $H$=3, similar to the feature in Fig.\ref{FIG2}b. However, in contrast with the meta-stable localization of $N$=4 electrons, opposite edge hole localization is not recovered as $H$ is increased to 4. This weaker aspect of hole localization is also observed more quantitatively in the power spectrum peak heights in Fig.\ref{FIG7}d which are weaker for both edges compared to Fig.\ref{FIG5}b.

In summary, we have investigated the Wigner crystal formation on the edges of periodic and finite zigzag graphene nanoribbons with different sizes using many-body configuration interaction calculations. The real space pair-correlation functions and power spectra reveal that Wigner crystallization occurs for ribbons above a critical width of 16 \mbox{\AA}. Across such distances, opposite edge localization also occurs although weaker than same edge localization. Further investigation of interedge correlations reveals an even-odd effect as a function of number of electrons. For odd number of electrons, Wigner localization is partially suppressed due degenerate many body state and electron sharing between two edges. Moreover, the analysis of addition energy spectrum shows the formation of meta-stable ground states for even number of electrons, leading to oscillatory behaviour consistent with classical calculations. Finally, we have shown that, although weaker than electron localization, a Wigner crystallization of holes is also possible at the edges of ZGNRs.

\section{Acknowledgment}
Authors acknowledge grant from the Scientific and Technological Research Council of Turkey (TUBITAK) under the project number 114F331. Modarresi thanks Dr. Pawel Potasz for fruitful discussions.

\section{Appendix: The mean field approach}
For a typical graphene quantum dot the general many-body equation may be simplified in the mean-field approach as, 

\begin{eqnarray}
H^{MF}= \sum_{p,\sigma} E_{p,\sigma} c_{p,\sigma}^{\dagger }c_{p,\sigma}+\sum_{p,q,r,s}\sum_{\sigma,\sigma^{'}} (\rho_{qr,\sigma^{'}}^{ZGNR}-\rho_{qr,\sigma^{'}}^{0}) \cr
(\left \langle pq \left | V \right | rs\right \rangle-\left \langle pq \left | V \right | rs\right \rangle \delta_{\sigma,\sigma^{'}}) c_{p,\sigma}^{\dagger } c_{q,\sigma^{'}}^{\dagger }c_{r,\sigma^{'}}c_{s,\sigma}
\end{eqnarray}

where the $\rho_{qr,\sigma^{'}}^{ZGNR}$ and $\rho_{qr,\sigma^{'}}^{0}$ are the electronic density for the ZGNR and bulk graphene, respectively. The above mean field Hamiltonian may be simplified to obtain a modified tight-binding Hamiltonian. The mean field introduces on-site and hopping like terms into the total Hamiltonian. We define the net electronic density for ZGNR as, $\rho_{qr,\sigma}=\rho_{qr,\sigma}^{ZGNR}-\rho_{qr,\sigma}^{0}$. The correction terms due to the the electron-electron interaction in the mean field sense are as,

\begin{multline}
H^{MF}(A,A)=v_{1111}\rho_{AA}+(2v_{1221}-v_{1212})\rho_{BB} \cr
+2v_{1331}\rho_{C_{A}C_{A}}+v_{1112}(\rho_{AB}+\rho_{BA})+2v_{1231}(\rho_{BX_{BA}}+\rho_{X_{BA}B}) \cr
+v_{1113}(\rho_{AC_{A}}+\rho_{C_{A}A})+(2v_{1223}-v_{1232}) \rho_{BX_{AB}}
\end{multline}

\begin{multline}
H^{MF}(A,B)=v_{1112}(\rho_{AA}+\rho_{BB}) \cr
+v_{1231}(2\rho_{X_{AB}X_{AB}}+2\rho_{X_{BA}X_{BA}}-\rho_{BX_{AB}}-\rho_{X_{BA}A}) \cr
+(2v_{1212}-v_{1221})\rho_{BA}+v_{1122}\rho_{AB}+v_{2213}(\rho_{X_{BA}B}+\rho_{AX_{AB}}) \cr
+(2v_{1232}-v_{1223})(\rho_{BX_{BA}}+\rho_{X_{AB}A})
\end{multline}

 \begin{multline}
H^{MF}(A,C_{A})=-v_{1331}(\rho_{C_{A}A}+\rho_{C_{A}B}) \cr
-v_{1231}(\rho_{B_{M_{A,C_{A}}}A}+\rho_{C_{A}BM_{{A,C_{A}}}})+v_{1113}(\rho_{AA}+\rho_{C_{A}C_{A}}) \cr
+(2v_{1223}-v_{1232})\rho_{BM_{{A,C_{A}}}BM_{{A,C_{A}}}}
 \end{multline}
 
where A and B are nearest-neighbor, A and $C_{A}$ are next-nearest-neighbor atoms, $X_{AB}$ is the nearest-neighbor of A and next-nearest-neighbor of B and $BM_{{A,B}}$ is the common nearest-neighbor of A and B atoms. The above equations include the direct and exchange terms for nearest and next-nearest neighbor atoms. The required parameters are tabulated in table I.

\begin{table}[ht]
\caption {The electron-electron interaction strength for graphene quantum dot.}
\begin{center}
\begin{tabular}{ |c|c|c| } 
 \hline
 Parameter  &   Value ($eV$)\\
 \toprule
 $V_{1111}$ & 16.5219\\ 
 $V_{1221}$ & 8.6396\\ 
 $V_{1331}$ & 5.3332\\ 
 $V_{1112}$ & 3.1574\\ 
 $V_{1231}$ & 1.7355\\
 $V_{1212}$ & 0.8729\\ 
 $V_{1122}$ & 0.8729\\ 
 $V_{2213}$ & 0.6061\\ 
 $V_{1113}$ & 0.3511\\ 
 $V_{1223}$ & 0.4094\\
 $V_{1232}$ & 0.6061\\
 \hline
\end{tabular}
\end{center}
\end{table}

The interaction strength in the dot is divided by dielectric constant $\kappa$ which is set to 6 for graphene quantum dot $v_{ijkl}=V_{pqrs}/\kappa$. By including the long-range Coulomb electron-electron interaction between A and D atoms, the mean field Hamiltonian is modified as,

 \begin{eqnarray}
H^{MF}(A,A)=\frac{2ke}{\kappa r_{AD}}\rho_{DD}
 \end{eqnarray}
 \begin{eqnarray}
H^{MF}(A,D)=\frac{-ke}{\kappa r_{AD}}\rho_{DA}
 \end{eqnarray}

where $k$ and $e$ are Coulomb's constant and electron elementary charge.

\section{References}
\bibliographystyle{unsrt}
\bibliographystyle{apsrev4-1}

\end{document}